\pdfoutput=1
\documentclass[aps,pra,10pt,twocolumn,showpacs,superscriptaddress,nofootinbib]{revtex4-1}

\usepackage[usenames,dvipsnames,svgnames]{pstricks}

\usepackage{newlfont}
\usepackage{amssymb}
\usepackage{amsfonts}
\usepackage{amsmath}
\usepackage{amsthm}
\usepackage{graphicx}

\usepackage{color}
\usepackage{bm}
\usepackage{times}

\begin{document}
\title{Non-commutative space: boon or bane for quantum engines and refrigerators}
\author{Pritam Chattopadhyay}
\email{pritam.cphys@gmail.com}
\affiliation{Cryptology and Security Research Unit, R.C. Bose Center for Cryptology and Security,\\
Indian Statistical Institute, Kolkata 700108, India}


\begin{abstract}
Various quantum systems are considered as the working substance for the analysis of quantum heat cycles and quantum refrigerators.  The ongoing technological challenge is how efficiently can a heat engine convert thermal energy to mechanical work. The seminal work of Carnot has proposed a fundamental upper limit--the Carnot limit on the efficiency of the heat engine.  However, the heat engines can be operated beyond the fundamental upper limit by exploiting non-equilibrium reservoirs. Here, the change in the space structure introduces the non-equilibrium effect. So, a question arises whether a change in the space structure can provide any boost for the quantum engines and refrigerators. The efficiency of the heat cycle and the coefficient of performance (COP) of the refrigerator cycles in the non-commutative space are analyzed here. The efficiency of the quantum heat engines gets a boost with the change in the space structure than the traditional quantum heat engine but the effectiveness of the non-commutative parameter is less for the efficiency compared to the COP of the refrigerator. There is a steep boost for the coefficient of performance of the refrigerator cycles for the non-commutative space harmonic oscillator compared to the harmonic oscillator. 
\end{abstract}

\maketitle

\section{Introduction}
In the early stage of quantum field theory, it was Heisenberg suggestion to use the non-commutative structure for space-time coordinates for small length scale. Synder \cite{syn} proposed a formalism in this area to control the divergence, which had troubled theories like quantum electrodynamics. To define the non-commutative space, we surrogate the space-time coordinate $x^i$ by their hermitian operators $\hat{x}^i$ of non-commutative $C^*$- algebra of space-time \cite{landi}. An example of physics in the non-commutative space-time is Yang--Mills theory on non-commutative torus \cite{connes}. Out of various versions, the most commonly studied version consists of replacing the standard version of commutation relations by their non-commutative version such as $[x^\mu, x^\nu]= i\theta^{\mu \nu}$, where $x^\mu, x^\nu$ are the canonical coordinates and $\theta^{\mu \nu}$  is considered as a constant antisymmetric tensor. Other interesting structures exist in the literature which leads to instances like minimal length and generalized version of Heisenberg's uncertainty relations. These are obtained when $\theta^{\mu \nu}$ is considered to be a function of the coordinates and momenta \cite{bbag,dey}.

Thermodynamics is a pre-eminent theory to evaluate the performance of the engines. It is one of the pillars of theoretical physics. Quantum thermodynamics deals with the exploration of the thermodynamic variables such as heat, temperature, entropy, energy and work for microscopic quantum systems and even for a single particle. The exploration of quantum thermodynamics involves the study of heat engines and refrigerators in microscopic regime \cite{skr,kolar,jro,rdo}. Heat engines exist in two forms one, is discrete and the other is continuous in nature. In a discrete group, we have two-stroke and the four-stroke engines and in the continuous group, we have a turbine. Various working models for the realization of the quantum engines have been proposed in previous works \cite{oab,ama,kzha,adec}. The pioneering work of Carnot suggested that the efficiency of the heat engine to convert thermal energy to mechanical work has a limit. The modern-day challenge is to develop a more efficient heat engine to convert thermal energy to mechanical work with the different working mediums. Theoretical studies suggest that the limit to the efficiency of the engine, i.e., the Carnot limit, can be surpassed by exploiting the non-equilibrium reservoirs. Now the question is can any working model in quantum regime exceed the Carnot efficiency and can it boost the Coefficient of performance (COP) of the refrigerator?

In this paper, we have proposed an approach to surpass the Carnot efficiency of the thermal machine based on the non-commutative space structure. Progress in this direction but with different approaches is shown in the works \cite{san1,san2,san3,pan12}. For our analysis, we utilize the latter version of non-commutative space-time where $\theta^{\mu \nu}$ is considered to be a function of the coordinates and momenta. Our prime motivation is to develop engine and refrigerator in non-commutative space where the working substance will be the perturbed harmonic oscillator in this space. We employ this harmonic oscillator in the Stirling and Otto cycle which is the working principle for different engines and refrigerator. We analyze all the stages of the cycle to compute the efficiency of this model. The outcomes are astonishing when compared with the results of the usual spaces. We always observe higher efficiency in non-commutative space than the usual spaces. Along with that, the most interesting observation is that the efficiency is more for this space structure than the commutative phase space when we switch on the non-commutative parameter but it decreases with the increase in the parameter. Whereas, with the increase in the non-commutative parameter, the COP rises correspondingly. This guides us to the possibility of using non-commutative systems for the exploration of quantum information processing to obtain better results. One immediate question that arises is whether the defined non-commutative system is accessible physically. The obvious answer is yes and it is shown in previous works \cite{afing,ipik}. The schematic analysis of the experimental model to access non-commutative space using optics is analyzed \cite{ipik}. Using the same methodology, one can think of modeling the heat engine of non-commutative space.

\section{Non-commutative harmonic oscillator (NHO)}

We initiate our discussion with the basic notion of squeezed state. We can obtain the squeezed states by applying the Glauber’s unitary displacement operator $D(\alpha)$ on the squeezed vacuum state \cite{mmn}. The mathematical form is $|\alpha,\xi\rangle = D(\alpha) A(\xi)|0\rangle,$  where $D(\alpha)=e^{(\alpha \alpha^{\dagger}- \alpha^{\ast} \alpha)}$ and $A(\xi)= e^{(\xi a^{\dagger}a^{\dagger}-\xi^{\ast}aa)}$. Here $\alpha$, $\xi$ are the displacement and squeezing parameters, respectively, and $A(\xi)$ is the unitary
squeezing operator. The ordering of the displacement and the squeezing operator are equivalent and it amounts to a change of parameter \cite{mmn}. We can also construct the squeezed state $|\alpha,\xi\rangle$ using the ladder operator.  It is obtained by performing the Holstein-Primakoff/Bogoliubov transformation on the squeezing operator \cite{mmn}.  It is defined as $ (a+\xi a^{\dagger})|\alpha,\xi\rangle = \alpha |\alpha,\xi\rangle$ where $a,\xi \, \epsilon \,\,\mathbb{C}$. The operators $a, a^{\dagger}$ are the bosonic annihilation and creation operators, i.e., $a^{\dagger} |n\rangle =\sqrt{k(n+1)}|n+1\rangle$ and $a|n\rangle= \sqrt{k(n)} |n-1\rangle$, where $k(n)$ is a general function which leads to different generalized models.   

The one-dimensional harmonic oscillator in the non-commutative space is defined as \cite{bbag,sdey,sghosh}
\begin{equation}\label{eq1}
H = \frac{P^2}{2m}+ \frac{1}{2} m\omega^2 X^2 - \hbar \omega \Big(\frac{1}{2}+\frac{\gamma}{4}\Big), 
\end{equation}
satisfying the relations
\begin{equation}\label{eq2}
[X,P]=i\hbar(1+\tilde{\gamma}P^2), \quad X=(1+\tilde{\gamma}p^2)x, \quad P=p.
\end{equation}
 
 Here $\gamma$ is a dimensionless constant and $\tilde{\gamma} =\gamma/(m\omega\hbar)$ has the dimension of inverse squared momentum. The observables $X,P$ representing the non-commutative space shown in ~\eqref{eq2} are expressed in terms of the standard canonical variables $x,p$ satisfying $[x,p]=i\hbar$. The Hamiltonian defined in ~\eqref{eq1} is non-hermitian with respect to the inner product.  The non-commutative Hamiltonian of the one-dimensional harmonic oscillator is derived from the standard one-dimensional harmonic oscillator which satisfies the condition $[X_i, P_j] = i \hbar (1 + \tilde{\gamma} P^2)$. The last term on the right-hand side appears during this transformation. The right side of the equation has a parameter $\gamma$, which represents the non-commutative parameter. We can tune the energy spectrum of the system by varying the non-commutative parameter.

We can construct the Hermitian counterpart of the defined non-Hermitian Hamiltonian shown in ~\eqref{eq1},  if we assume the Hamiltonian $H$ to be a pseudo-Hermitian, i.e., the non-Hermitian Hamiltonian $H$ and the Hermitian Hamiltonian $h$ are interlinked by a similarity transformation $h=\mu H \mu^{-1}$. Here $\mu \mu^{\dagger}$ is the positive definite operator, and it plays the role of the metric. Now, we can represent the eigenstates of the Hamiltonians $H$ and $h$ as $|\phi\rangle$ and $|\varphi\rangle$, respectively. They are related as $|\phi\rangle = \mu^{-1} |\varphi\rangle$. The Dyson map $\mu$ takes the form $\mu = (1+\tilde{\gamma}p^2)^{-1/2}$. The Dyson map $\mu$ is defined by the same set of operators as that of the Hamiltonian. The Dyson map can be expressed in the general form as shown in previous works \cite{dey123,dos123} with some assumption. Following the same methodology, the relation between the Dyson map $\mu$ and the Hamiltonian parameters can be developed which in turn satisfies the commutation relations $[x_i , x_j] = i \theta_{ij}$, where $\theta_{ij}$  is considered as a constant antisymmetric tensor. Considering this metric, we can compute the Hamiltonian $h$ as

\begin{eqnarray}\label{eq3}\nonumber
h & =& \mu H \mu^{-1}\\ \nonumber
& = & \frac{p^2}{2m}+\frac{1}{2}\omega^2 x^2 + \frac{\omega \mu}{4\hbar}\Big(p^2x^2+x^2p^2+2xp^2x\Big)\\
& - & \hbar \omega \Big(\frac{1}{2}+\frac{\mu}{4} \Big)+ \mathcal{O}(\mu^2).
\end{eqnarray}

Now we will use the perturbation treatment to decompose the Hamiltonian~\eqref{eq3} into $h = h_0 + h_1$, where $h_0$ is taken to be the standard harmonic oscillator and $h_1$ as the perturbation part. Using the perturbation theory, the  energy eigenvalues of $H$ and $h$ evolves to
\begin{equation}\label{eq4}
E_n = \hbar \omega(A n+B n^2) + \mathcal{O}(\mu^2),
\end{equation}
where $A= (1+\mu/2)$ and $B=\mu/2$. The corresponding eigenstates of the system are
 
\begin{eqnarray}\label{eq5}\nonumber
|\psi_n\rangle &= & |n\rangle - \frac{\gamma}{16}\sqrt{(n-3)^{(4)}} |n-4\rangle \\
& + & \frac{\gamma}{16}\sqrt{(n+1)^{(4)}} |n+4\rangle + \mathcal{O}(\mu^2),
\end{eqnarray}
where $P = \Pi_{k=0}^{n-1}{(P+k)}$ represents  the Pochhammer symbol with the raising factorial. Now we have all the prerequisites for the analysis of the thermodynamic cycles in this space structures.

\section{Quantum heat engines of non-commutative space}
The canonical partition function for the defined Hamiltonian~\eqref{eq1} can be evaluated with the help of its corresponding eigenvalue to 
\begin{eqnarray} \label{eqn1} \nonumber
Z &= & \sum_n e^{-\beta E_n } \\ 
 & = & \frac{e^{\frac{\beta (2+\mu)^2 \omega}{8\mu}} \sqrt{\frac{\pi}{2}} Erfc \Big[\frac{\beta(2+\mu)\omega}{2\sqrt{2\beta \mu \omega}}\Big]}{\sqrt{\beta \mu \omega}},
\end{eqnarray}
subjected to the condition that $Re[\beta \omega \mu] > 0$.
 We are neglecting the higher order terms because they tend to zero for the higher order.  Erfc is the complementary error function, it is defined as $erfc (x) =\frac{\Gamma(1/2, x^2)}{\sqrt{\pi}} $, where $\Gamma \left( n, x \right)$ is the incomplete gamma function. It is expressed as $\Gamma \left( n, x \right) = (n-1)! e^x \sum_{k=0}^{n-1} \frac{x^k}{k!}$. We will now be able to evaluate all the thermodynamic variables in terms of the established partition function of the considered system for the analysis of the engine model.

\subsection{Quantum stirling heat cycle}

\begin{figure} [h]
\centering
  \includegraphics[width=1.0\linewidth]{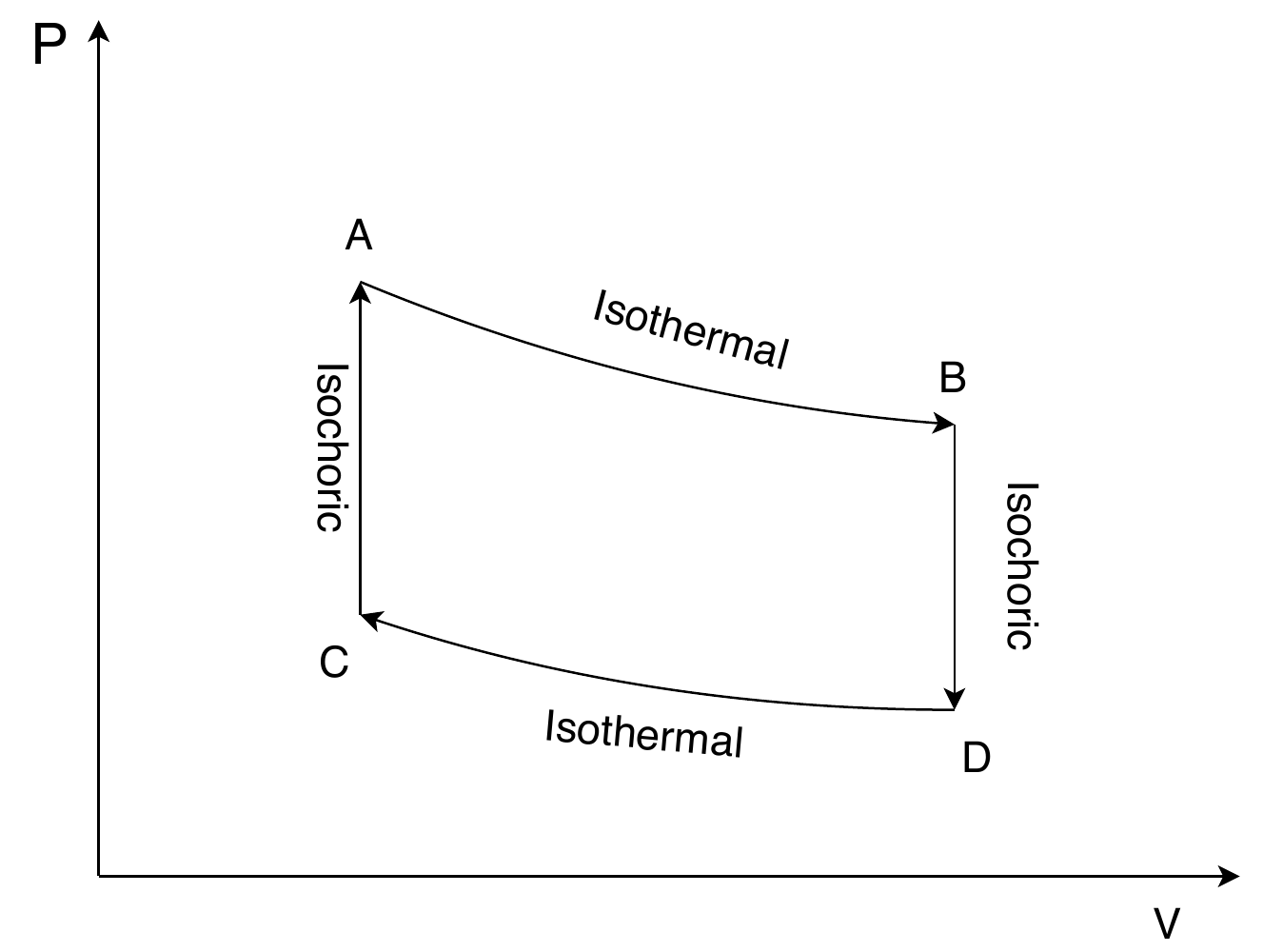}
\caption{Schematic diagram for the four stages of the Stirling cycle.}
\label{Fig1}
\end{figure}

We will analyze the well-known \textit{Stirling cycle} with non-commutating harmonic oscillator as the working substance. The four stages of the Stirling cycle \cite{gsa,blic,tho,pc,pc1} (Fig. (\ref{Fig1})) is as follows: 

(i) The first step of the cycle is the \textit{isothermal (A$\rightarrow$B)} process. In this process, the working substance will be kept in contact with a heat bath of temperature $T_h$. The system stays in thermal equilibrium with the heat bath throughout every instant of time. The energy spectrum $E_n$ and the internal energy $U$ are changed as a result of the slow change in the working substance, i.e., the changes that take place in Hamiltonian during the execution of this phase. So, heat is absorbed from the bath in this phase. The heat exchange during this phase of the cycle is
\begin{eqnarray} \label{eqn2}
Q_{AB} = U_B - U_A + k_B T_h ln Z_B - k_B T_h ln Z_A,
\end{eqnarray}
where $k_B$ is the Boltzmann constant.

(ii) The second phase of the cycle is the \textit{isochoric (B$\rightarrow$C)} process. During this process, the system undergoes an isochoric heat exchange. The system is connected to a bath with lower temperature $T_c$, so heat is released. The heat exchange for this process is expressed as 

\begin{eqnarray} \label{eqn3}
Q_{BC}= U_C -U_B.
\end{eqnarray}

(iii) The third phase is an \textit{isothermal (C$\rightarrow$D)} process. During this phase of the cycle, the working substance is kept in contact with a bath of lower temperature $T_c$. Similar to phase one, the system is in thermal equilibrium with the bath. In this process, heat is released. The heat exchange during this stage of the cycle is 

\begin{eqnarray} \label{eqn4}
Q_{CD} = U_D - U_C + k_B T_c ln Z_D - k_B T_c ln Z_C.
\end{eqnarray}

(iv) The last stage of the cycle is the \textit{isochoric (D$\rightarrow$A)} process. The system is connected back to the bath with a higher temperature $T_h$. The heat exchange for the last stage of the cycle is expressed as
\begin{eqnarray} \label{eqn5}
Q_{DA} = U_A-U_D.
\end{eqnarray}

For all the phases, the internal energy of the system can be evaluated using the partition function as $U= -\frac{\partial ln Z}{\partial \beta}$. The different form of the partition function ($Z_A, Z_B, Z_C, Z_D$)  arises due to the changes that occur in the Hamiltonian of the system during the different phases of the cycle.  The total work done is $W_{tot} = Q_{AB} + Q_{BC} + Q_{CD} + Q_{DA}$. The efficiency of the Stirling heat cycle is expressed as 

\begin{eqnarray} \label{eqn6}
\eta_{Stir}= 1 + \frac{Q_{BC} + Q_{CD}}{Q_{DA} + Q_{AB}}.
\end{eqnarray}

\begin{figure}[h]
  \centering
  \includegraphics[width=1.0\linewidth]{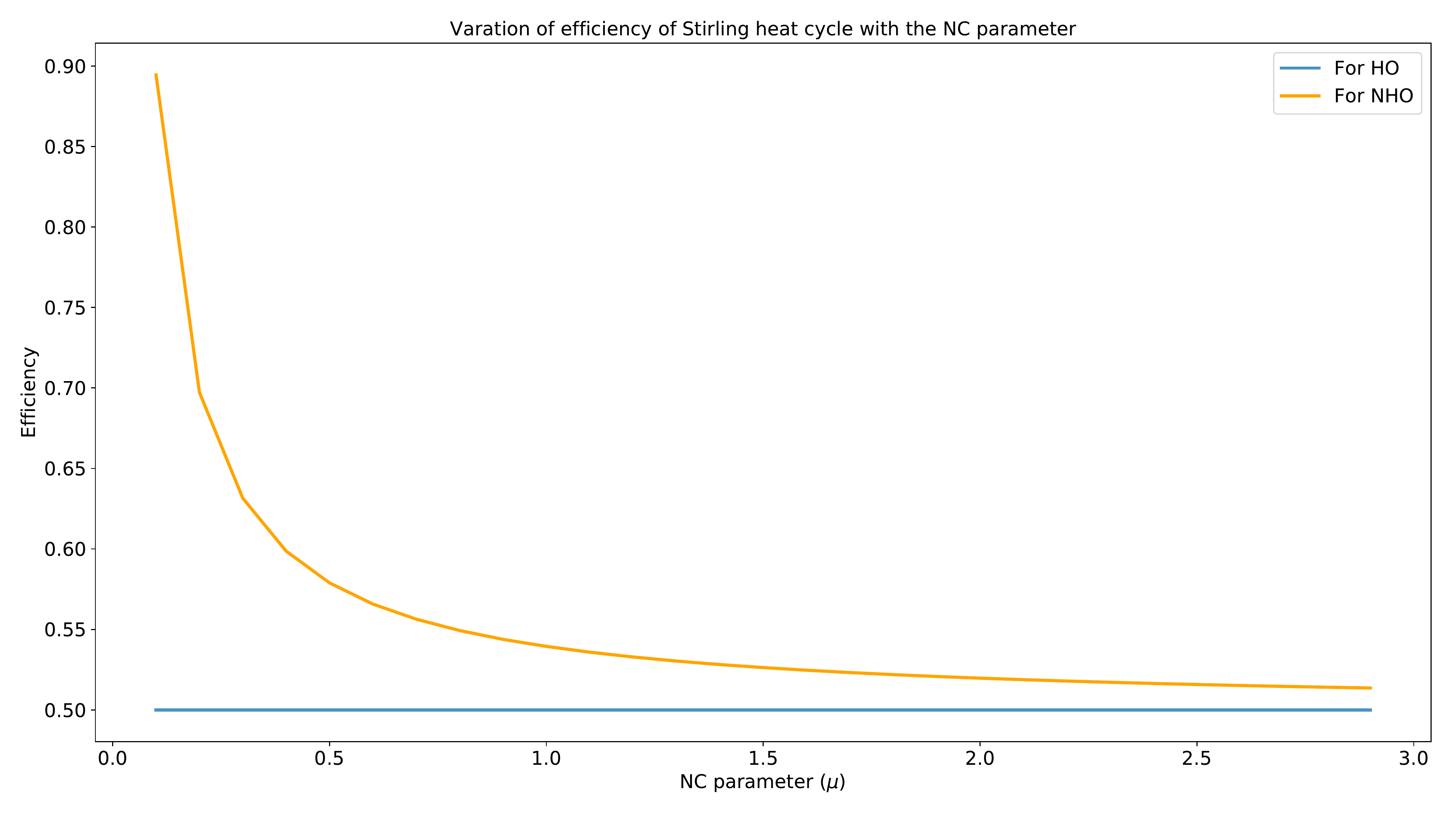}
\caption{(Color Online) The variation of the efficiency of the Stirling cycle for the HO and NHO. The temperature of the hot bath and the cold bath is $T_h= 20K$ and $T_c = 10K$ respectively. The yellow and the blue solid line is the variation of the COP of NHO and HO with NC parameter, respectively.}
\label{Fig3}
\end{figure}

\subsection{Stirling refrigerator cycle}

If we reverse the cycle, we will have a Stirling refrigerator \cite{xlh}. Following the same methodology, as done above, we can analyze all the four stages of the cycle. 

(i) The first phase is the \textit{isothermal} process. In this process, the system is paired to a cold bath at temperature $T_c$. This is just the reverse of the first phase of the heat cycle. The entropy of the system changes during this process. The heat absorbed is 

\begin{equation}
Q_{AB} = T_c \Delta S.
\end{equation}

(ii) The second stage is the \textit{isochoric} process. During this stage of the cycle, the temperature  of  the  system  increases when connected to $T_h$ from $T_c$. The mean internal energy of the system changes during this phase of the working cycle. The heat gain for this phase is
\begin{equation}
Q_{BC} = U_C - U_B.
\end{equation}
(iii) This phase is an \textit{isothermal} process. During this stage of the cycle, the system is bridged with the hot reservoir with a temperature $T_h$. Heat is rejected from the system and is described as

\begin{equation}
Q_{CD} = T_h \Delta S.
\end{equation} 

(iv) The last stage is an \textit{isochoric} process. The system is reverted to the cold reservoir $T_c$ from the hot reservoir $T_h $, which leads to a decrease in the internal energy of the system. The amount of heat released is
\begin{equation}
Q_{DA} = U_A - U_D.
\end{equation}
The entropy of the system can be evaluated from the partition function of the system as $S = \ln Z + \beta U$. The internal energy can be evaluated as shown while we have analyzed the heat cycle.
The COP of the Stirling refrigerator cycle is expressed as,
\begin{equation}
COP_{Stir} = \frac{Q_{AB} + Q_{BC}}{W_{T}},
\label{St-cop}
\end{equation}
where $W_{T}= Q_{AB} + Q_{BC} + Q_{CD} + Q_{DA}$ is the total work done on the system.

\begin{figure} [h]
\centering
  \includegraphics[width=1.0\linewidth]{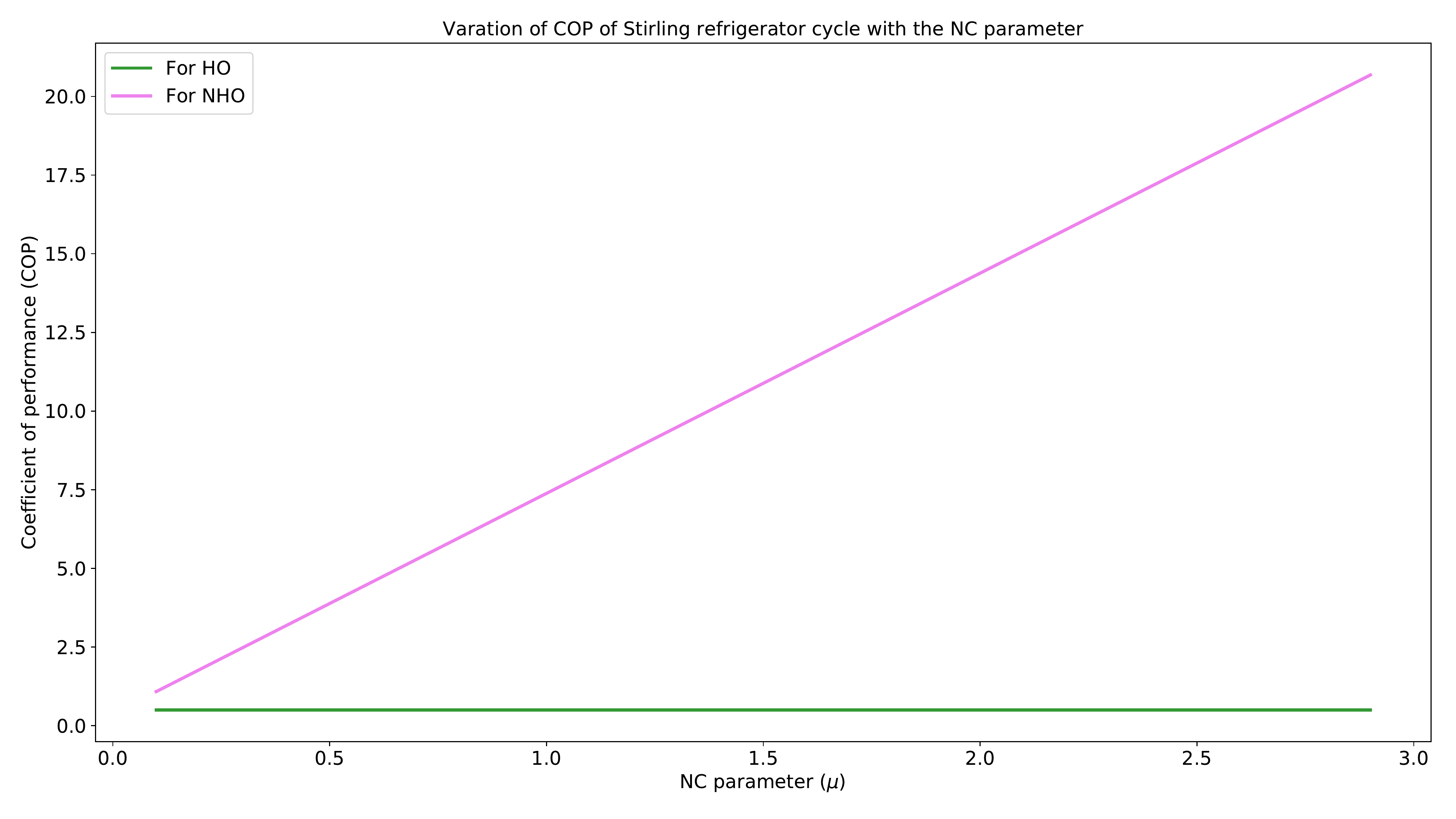}
\caption{(Color Online) The variation of the COP of the Stirling refrigerator cycle for the Harmonic Oscillator (HO) and NHO. The temperature of the hot bath and the cold bath is $T_h= 20K$ and $T_c = 10K$ respectively. The violet and the green solid line is the variation of the COP of NHO and HO with NC parameter, respectively.}
\label{Fig2}
\end{figure}

We can visualize the growth in the COP for the Stirling cycle for NHO due to the non-commutating parameter. Whereas for the case of HO the COP remains constant as it is independent of this parameter.  For $\omega>\omega'$ and $\beta_h < \beta_c$ we encounter a steep rise in the COP with respect to the NC parameter as shown in Fig. (\ref{Fig2}).  In Fig. (\ref{Fig3}), the efficiency of the engine decreases with the increase in the NC parameter. So the non-commutative is less effective when the NC parameter is high for the engine model that we have considered for our analysis. For the generic statement of the less effectiveness of NC parameter on the engine model, we have to explore other cycles which is an open area for exploration. 

The maximum attainable efficiency of the heat engine by the standard harmonic oscillator has been plotted as a reference point for the analysis of the advantage due to the NC space. Now, due to the change in the space structure, the standard Hamiltonian changes to the non-commutative harmonic oscillator by applying the transformation from the commutative space to the non-commutative space. So, to compare the advantage of the change introduced by the non-commutative space, we have considered the standard harmonic oscillator as a reference point. For the Stirling cycle, we can visualize the advantage in Fig. (\ref{Fig3}) and Fig. (\ref{Fig2}).

\subsection{Quantum Otto cycle}

\begin{figure}
  \centering
  \includegraphics[width=1.0\linewidth]{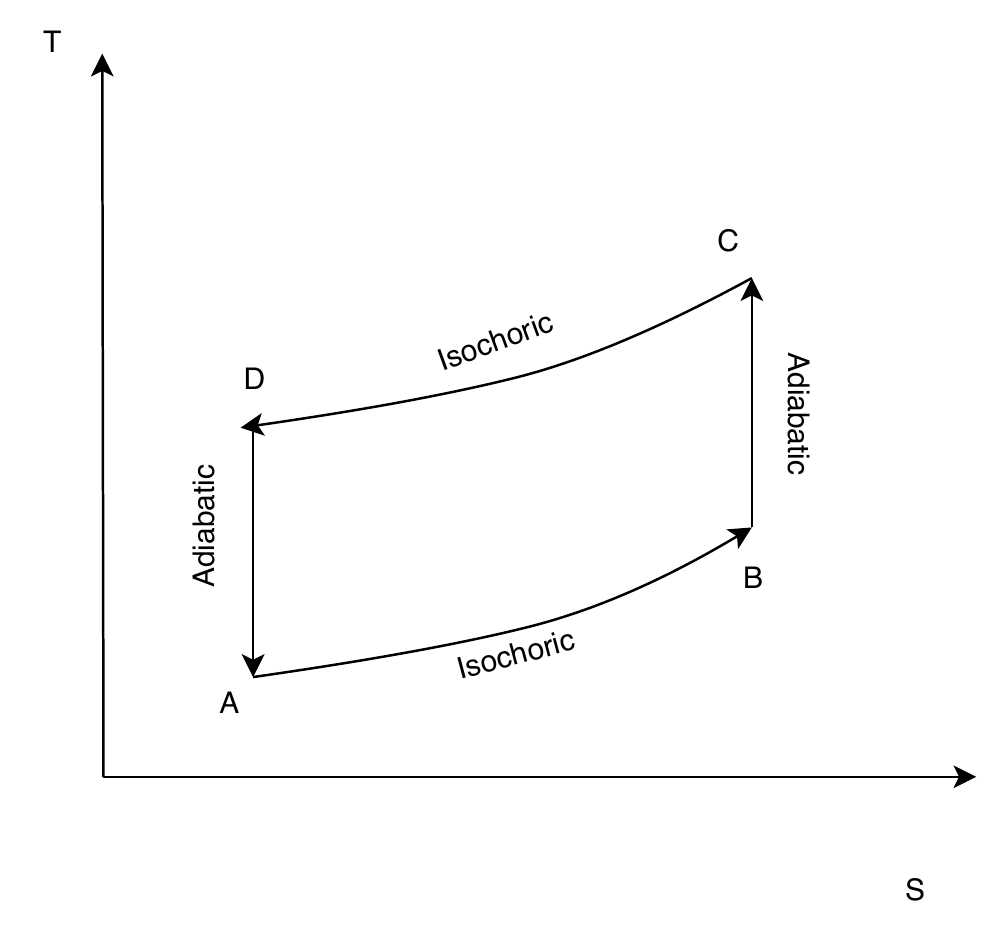}
\caption{Schematic diagram for the four stages of the Otto cycle.}
\label{Fig4}
\end{figure}

Now we will study the quantum \textit{Otto refrigerator cycle} \cite{htq,kos,long} with non-commutative space harmonic oscillator as the working substance. The four phases of the Otto refrigerator  (Fig. (\ref{Fig4})) are described as follows: 

(i) The first phase of the cycle is an \textit{isochoric (A$\rightarrow$B)} process. During this process, the system is coupled to a cold reservoir at a temperature $T_C$ while the Hamiltonian remains constant. The heat absorbed from the reservoir during this process is

\begin{equation} \label{eqn7}
Q_{cold} = \sum_n E_n^{cold} (P_n^{hot}-P_n^{cold}),
\end{equation}
where $ P_n^{cold}= \frac{\exp(-\beta E_{n})}{Z}|_{\beta=\beta_{cold},\omega=\omega'}$ 
and\\ $P_{n}^{hot}= \frac{\exp(-\beta E_{n})}{Z}|_{\beta=\beta_{hot},\omega=\omega}$ represents the occupation probabilities of the system  in  the  nth  eigenstate and $E_n^{cold} =E_n$ for $\omega=\omega'$. 

(ii) The second stage of the cycle is an \textit{adiabatic (B$\rightarrow$C)} process. During this phase of the cycle, the entropy of the system is conserved. Throughout the evolution of the system in this phase, the occupation distribution remains invariant. 

(iii) The third stage of the cycle is an \textit{isochoric (C$\rightarrow$D)} heating process. In this process, the system is connected to a hot reservoir at temperature $T_H$. The heat rejected to the hot reservoir in this phase is
\begin{equation} \label{eqn8}
Q_{hot} = \sum_{n} E_{n}^{hot} (P_{n}^{cold} - P_{n}^{hot}),
\end{equation}
where $E_{n}^{hot}= E_n$ for $\omega=\omega$.

(iv) The last stage of the cycle is an \textit{adiabatic (D$\rightarrow$A)} process. In this process, the system changes quasi-statically while the entropy of the system remains constant during the execution of this phase. The total work done on the cycle can be evaluated as, $W_{total} = Q_{hot} + Q_{cold}$. The COP for the Otto refrigerator is defined as the ratio of the amount of heat removed form the cold reservoir to the net amount of work done on the system under analysis. It is represented as 
\begin{equation} \label{eqn9}
\Xi_{Otto} = \frac{Q_{cold}}{|W_{total}|}.
\end{equation}

\begin{figure}
\includegraphics[width=1.0\linewidth]{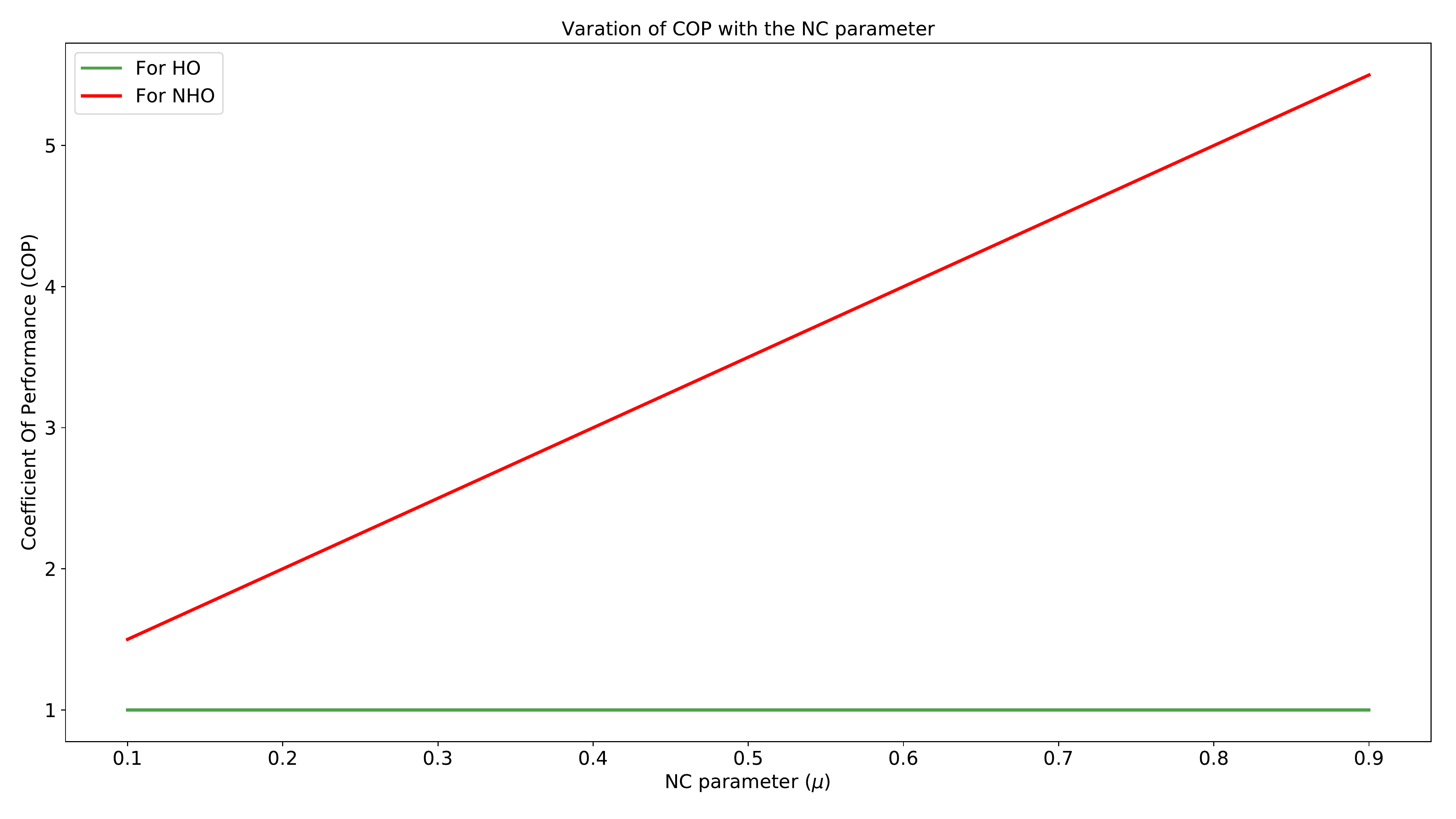}
\caption{(Color Online) The plot depicts the evaluation of the COP of the Otto refrigerator for the HO and NHO. The temperature of the hot and the cold bath is $T_h= 20K$ and $T_c = 10K$ respectively. The red and the green solid line is the variation of the COP of NHO and HO with NC parameter, respectively.}
\label{Fig5}
\end{figure}

In the case of the Otto refrigerator, we have encountered a similar effect as in the case of Stirling cycle. We have detected the growth in the COP for the Otto cycle for NHO due to the non-commutating parameter shown in Fig. (\ref{Fig5}). But in the case of HO the COP remains constant. A steep rise in the coefficient of the NC parameter occurs for $\omega>\omega'$ and $\beta_h < \beta_c$ and this gives rise to this phenomenon. Following a similar pattern as done during the analysis of the Stirling cycle, we have considered the maximum attainable efficiency of the heat engine by the standard harmonic oscillator has been plotted as a reference point for the analysis of the advantage due to the NC space. For the Otto cycle, we can visualize the advantage in Fig. (\ref{Fig5}).

The immediate question that pokes our intuition is whether it is feasible with quantum technology we have? The answer to this is yes. One can analysis the non-commutative space effect using optical setup and measure the effect of the non-commutative parameter as shown in \cite{ipik}. They have provided a schematic representation of the experimental setup for the following analysis. Following the same methodology, we can develop the setup for  the analysis of different thermodynamic cycles. For experimental realization of the cycle in non-commutative space, one should keep in mind the basic ingredients that are required for the analysis. One of which is the availability of thermal heat bath for the different processes. The second one is about the measurement of the work performed during the different phases of the cycle, as in the case of the Otto cycle, the phases are two adiabatic processes. One of which expands the working medium and the other compresses the working medium. Third one is maintaining the thermal equilibrium during the thermalization processes. The experimental analysis of the engines and refrigerators of NC space is an open area to explore.

\section{Discussion and conclusion}
The non-commutative harmonic oscillator outperforms the harmonic oscillator in terms of the COP for the Stirling cycle and the Otto refrigerator. The contribution for this is provided by the non-commutative space parameter. So, we can infer that non-commutative is a boon for the refrigerators if considered for the growth of the COP. Whereas the NC parameter is less effective for a boost in the efficiency of the heat cycle for the considered model. 

For the appropriate implementation of Otto refrigerator, it requires a slow implementation of the adiabatic processes so that we can maintain no further coherence generation on the eigenstates of the non-commutative space Hamiltonian. If there is any change, then the mean population will also change. To achieve thermal equilibrium with the reservoir, the system must spend a long time during the thermalization processes. The non-linearity that is generated in the Hamiltonian appears due to the non-commutative parameter which requires some energy cost for the implementation of the cycle.

This model can result to a better resource in the applicable areas \cite{mhub} of quantum theory which needs further analysis. This model can be used for the analysis of the coupled working medium as shown in previous works \cite{tzh,jwa,gth}. It can also be utilized for exploring the non-Markovian reservoirs in NC space. We have analyzed two thermodynamic cycles. One can analyze the different existing reversible cycles and irreversible cycles. It will be interesting to explore the effect of NC parameters in the irreversible cycles and on the quantum phase transition. 

For the analysis of our work, we have used one of the existing models of the non-commutative space. To make the generic statement of the gain in COP for different cycles, one has to explore all the existing models in the non-commutative space. This is an open area to explore. One can also analyze the effect of the NC parameter on the different thermodynamic variables. We can analyze this model from uncertainty viewpoint \cite{pc,pc1,pc3} to reduce the cost for the analysis of the cycles. Along with that, it will be interesting to explore the experimental realization of these cycles in NC space.

\section*{Acknowledgment}
The author gratefully acknowledges for the useful discussions and suggestions from Dr. Goutam Paul, Associate Prof. of Cryptology and Security Research Unit, R.C. Bose Center for Cryptology and Security, at Indian Statistical Institute, Kolkata and Mr. Tanmoy Pandit of Hebrew University of Jerusalem, Jerusalem, Israel. We thank the reviewer for constructive comments.


\bibliographystyle{h-physrev4}

\begin{thebibliography}{99}

\bibitem{syn}Snyder, Hartland S. ``The electromagnetic field in quantized space-time." Physical Review 72.1 (1947): 68.

\bibitem{landi}G. Landi, ``An Introduction to Noncommutative Spaces and their Geometries” (Springer-Verlag, 1997).

\bibitem{connes}A. Connes and M.A. Rieffel, ``Yang-Mills for Noncommutative Two-Tori”, Contemp.Math.62 (1987) 237.

\bibitem{bbag}B. Bagchi and A. Fring,  ``Minimal length in quantum mechanics and non-Hermitian Hamiltonian systems",  Phys. Lett. A373, 4307–4310 (2009).

\bibitem{dey}Dey, Sanjib, and Véronique Hussin. ``Entangled squeezed states in noncommutative spaces with minimal length uncertainty relations." Physical Review D 91.12 (2015): 124017.

\bibitem{skr}Skrzypczyk, Paul, Anthony J. Short, and Sandu Popescu. ``Work extraction and thermodynamics for individual quantum systems." Nature communications 5 (2014): 4185.

\bibitem{kolar}Kol\'a\v{r}, Michal, et al. ``Quantum bath refrigeration towards absolute zero: Challenging the unattainability principle." Physical review letters 109.9 (2012): 090601.

\bibitem{jro}J. Ronagel, O. Abah, F. Schmidt-Kaler, K. Singer, and E. Lutz, ``Nanoscale Heat Engine Beyond the Carnot Limit", Phys. Rev. Lett. 112, 030602 (2014).

\bibitem{rdo}R. Dorner, S. R. Clark, L. Heaney, R. Fazio, J. Goold, and V. Vedral, ``Extracting Quantum Work Statistics and Fluctuation Theorems by Single-Qubit Interferometry", Phys. Rev. Lett. 110, 230601 (2013).

\bibitem{oab}O. Abah, J. Rossnagel, G. Jacob, S. Deffner, F. Schmidt Kaler, K. Singer, and E. Lutz, ``Single-Ion Heat Engine at Maximum Power", Phys. Rev. Lett. 109, 203006 (2012).

\bibitem{ama}A. Mari and J. Eisert, ``Cooling by Heating: Very Hot Thermal Light Can Significantly Cool Quantum Systems", Phys. Rev. Lett. 108, 120602 (2012).

\bibitem{kzha}K. Zhang, F. Bariani, and P. Meystre, ``Quantum Opto-mechanical Heat Engine", Phys. Rev. Lett. 112, 150602 (2014).

\bibitem{adec}A. Dechant, N. Kiesel, and E. Lutz, ``All-Optical Nano-mechanical Heat Engine", Phys. Rev. Lett. 114, 183602 (2015).

\bibitem{san1}Santos, Jonas FG, and Alex E. Bernardini. ``Quantum engines and the range of the second law of thermodynamics in the noncommutative phase-space." The European Physical Journal Plus 132.6 (2017): 260.

\bibitem{san2}Santos, Jonas FG. ``Noncommutative phase-space effects in thermal diffusion of Gaussian states." Journal of Physics A: Mathematical and Theoretical 52.40 (2019): 405306.

\bibitem{san3}Santos, Jonas FG. ``Heat flow and noncommutative quantum mechanics in phase-space." arXiv preprint arXiv:1912.11884 (2019).

\bibitem{pan12}Pandit, Tanmoy, Pritam Chattopadhyay, and Goutam Paul. ``Non-commutative space engine: a boost to thermodynamic processes." arXiv preprint arXiv:1911.13105 (2019).

\bibitem{afing} A. Fring, L. Gouba and F. G. Scholtz, ``Strings from position-dependent noncommutativity", J. Phys. A: Math. Theor. 43, 345401 (2010).

\bibitem{ipik}I.  Pikovski  et  al.,  ``Probing  Planck-scale  physics  with quantum optics",  Nature Phys. 8, 393–397 (2012).

\bibitem{mmn}M. M. Nieto and D. R. Truax, ``Squeezed states for general systems", Phys. Rev. Lett. 71, 2843 (1993).

\bibitem{sdey} S.  Dey  and  A.  Fring,   ``Squeezed  coherent  states  for  noncommutative  spaces  with  minimal length uncertainty relations",  Phys. Rev. D86, 064038 (2012).

\bibitem{sghosh}S. Ghosh and P. Roy, ``Stringy coherent states inspired by generalized uncertainty principle", Phys. Lett. B 711, 423–427 (2012).

\bibitem{dey123}Dey, Sanjib, Andreas Fring, and Thilagarajah Mathanaranjan. ``Non-Hermitian systems of Euclidean Lie algebraic type with real energy spectra." Annals of Physics 346 (2014): 28-41.

\bibitem{dos123}dos Santos, J. F. G., et. al. ``Non-Hermitian noncommutative quantum mechanics." The European Physical Journal Plus 134.7 (2019): 332.

\bibitem{gsa}  G.  S.  Agarwal  and  S.  Chaturvedi, ``Quantum dynamical framework for Brownian heat engines," Phys.  Rev.  E 88,012130 (2013).

\bibitem{blic}Blickle, Valentin, and Clemens Bechinger. ``Realization of a micrometre-sized stochastic heat engine." Nature Physics 8.2 (2012): 143.

\bibitem{tho}Thomas, George, Debmalya Das, and Sibasish Ghosh. ``Quantum heat engine based on level degeneracy." Physical Review E 100.1 (2019): 012123.

\bibitem{pc}Chattopadhyay, Pritam, Ayan Mitra, and Goutam Paul. ``Bound on efficiency of heat engine from uncertainty relation viewpoint." arXiv preprint arXiv:1908.06804 (2019).

\bibitem{pc1}Chattopadhyay, Pritam, and Goutam Paul. ``Relativistic quantum heat engine from uncertainty relation standpoint." Scientific reports 9.1 (2019): 1-12.

\bibitem{xlh}X-L. Huang, X-Y. Niu, X-M. Xiu,  and X-X. Yi, ``Quantum stirling heat engine and refrigerator with single and coupled spin systems,” The European Physical Journal D 68, 32 (2014).

\bibitem{htq} H. T. Quan, Y-x. Liu, C. P. Sun,  and F. Nori, ``Quantum thermodynamic cycles and quantum heat engines," Phys. Rev. E 76,031105 (2007).

\bibitem{kos}Kosloff, Ronnie, and Yair Rezek. ``The quantum harmonic Otto cycle." Entropy 19.4 (2017): 136.

\bibitem{long}Long, Rui, and Wei Liu. ``Performance of quantum Otto refrigerators with squeezing." Physical Review E 91.6 (2015): 062137.

\bibitem{mhub}M. Huber,M. Perarnau-Llobet, K. V Hovhannisyan, P. Skrzypczyk, C. Klckl, N. Brunner,  and A. Acn, ``Thermodynamic cost of creating correlations," New J. Phys.17, 065008 (2015).

\bibitem{tzh}T. Zhang, W-T. Liu, P-X. Chen,  and C-Z. Li, ``Four-level entangled quantum heat engines," Phys. Rev. A 75, 062102 (2007).

\bibitem{jwa} J. Wang, Z. Ye, Y. Lai, W. Li,   and J. He, ``Efficiency at maximum power of a quantum heat engine based on two coupled oscillators," Phys. Rev. E 91, 062134 (2015).

\bibitem{gth} G. Thomas  and R. S. Johal, ``Coupled  quantum  otto  cycle," Phys. Rev. E 83, 031135 (2011).

\bibitem{pc3}Chattopadhyay, Pritam, Ayan Mitra, and Goutam Paul. ``Probing Uncertainty Relations in Non-Commutative Space." Int J Theor Phys (2019) 58: 2619.





\end{thebibliography}



\end{document}